\documentclass[10pt,tightenlines,eqsecnum,floats,aps,amsmath,amssymb,nofootinbib,prd,showpacs]{revtex4}
\usepackage{amssymb}
\usepackage{stmaryrd}
\usepackage{amsmath}
\usepackage{amsfonts}
\usepackage{mathrsfs}

\usepackage{amsmath,amssymb,amsfonts}
\usepackage{graphicx}

\def\be{\begin{equation}}
\def\ee{\end{equation}}
\def\ba{\begin{eqnarray}}
\def\ea{\end{eqnarray}}

\newcommand{\R}{\mathcal {R}} 

\begin{document}

\title{Extension of Loop Quantum Gravity to Metric Theories beyond General Relativity}

\author{Yongge Ma}\email{mayg@bnu.edu.cn}

\address{Department of Physics, Beijing Normal University,
Beijing 100875, China}

\begin{abstract}
The successful background-independent quantization of Loop Quantum Gravity relies on the key observation that
classical General Relativity can be cast into the connection-dynamical formalism with the structure group of $SU(2)$. Due to this
particular formalism, Loop Quantum Gravity was generally considered as a quantization scheme that applies only to General Relativity.
However, we will show that the nonperturbative quantization procedure of Loop Quantum Gravity can be extended to a rather general
class of metric theories of gravity, which have received increased attention recently due to
motivations coming form cosmology and astrophysics. In particular, we will first introduce how to reformulate the 4-dimensional metric
$f(R)$ theories of gravity, as well as Brans-Dicke theory, into connection-dynamical formalism with real
$SU(2)$ connections as configuration variables. Through these formalisms, we then outline the nonpertubative
canonical quantization of the $f(R)$ theories and Brans-Dicke theory by extending the loop quantization scheme
of General Relativity.
\end{abstract}


\maketitle

\section{Introduction}
In the recent 25 years, loop quantum gravity(LQG), a background
independent approach to quantize general relativity (GR), has been
widely investigated \cite{Ro04,Th07,As04,Ma07}. It is remarkable
that, as a non-renormalizable theory, GR can be non-perturbatively
quantized by the loop quantization procedure. This
background-independent quantization relies on the key observation
that classical GR can be cast into the connection-dynamical
formalism with the structure group of $SU(2)$\cite{As86,Ba}. Thus one is naturally led to ask the following questions. Does the non-perturbative quantization scheme of LQG apply only to GR? What is the applicable scope of LQG? On this Loops 11 Conference, we just learned from the talks by Thiemann et al that LQG is applicable to GR in arbitrary dimensions, since higher dimensional GR could also be cast into connection dynamics with certain compact structure group. We are going to answer the question: whether GR is a unique relativistic theory of gravity with the desired connection-dynamical character?

In fact, modified
gravity theories have recently received increased attention due to
motivations coming form cosmology and astrophysics. A series of
independent observations, including type Ia supernova, weak lens,
cosmic microwave background anisotropy, baryon oscillation, etc,
implied that our universe is currently undergoing a period of
accelerated expansion\cite{Fr08}. This result has carried the "dark energy" problem in GR. A large number of phenomenological models for dark energy have been proposed, such as quintessence, phantom, Chaplygin gas, K-essence etc.
    In all these models, certain scalar fields are added by hand, and their origins are not
  understood. Moreover, there are indirect evidences suggesting that the bulk of the matter of the universe is invisible or dark (see e.g. \cite{FT}). The strongest evidence comes from the rotational velocity of the isolated stars or hydrogen cloud on the outskirts of galaxies. Although a few candidates for dark matter are proposed, such as sterile neutrinos, neutralinos etc, none of them are satisfying. To explain the accelerated expansion of the universe, as well as dark matter, from
  fundamental physics is now a great challenge. It is reasonable to consider the possibility that GR is not a valid theory of gravity on a galactic or cosmological scale. Historically, Einstein's GR is the simplest relativistic theory of gravity with correct Newtonian limit. It is worth pursuing all alternatives, which provide a high chance to new physics.

  A large variety of models of $f(\mathcal{R})$ modified gravity have been proposed to explain the "dark universe" without recourse to dark energy and dark matter \cite{So}. The action of metric $f(\mathcal{R})$ theories reads:
\be
    S[g]=\frac12\int
d^4x\sqrt{-g}f(\mathcal {R}) \label{action0}
\ee
where $f$ is a general function of the scalar curvature $\mathcal{R}$, and we set $8\pi G=1$.
Besides $f(\R)$ theories, a well-known competing
relativistic theory of gravity was proposed by Brans and Dicke in
1961 \cite{BD}, which is apparently compatible with Mach's
principle. To represent a varying "gravitational constant", a scalar
field is non-minimally coupled to the metric as
\be
 S[g,\phi]=\int
d^4x\sqrt{-g}(\phi\mathcal{R}-\frac{\omega}{\phi}(\partial_\mu\phi)\partial^\mu\phi).\label{BD}
\ee
The
scalar field in Brans-Dicke theory (BDT) of gravity is then expected to account for "dark
energy". It can naturally lead to cosmological acceleration if certain potential term of the scalar is added in the original action (\ref{BD}). It is also possible to account for the dark matter problem. Furthermore, a large part of the non-trivial tests on
gravity theory is related to Einstein's equivalence principle (EEP)
\cite{will}, which contains the following three elements. (i) The trajectory of a freely falling "test" body is independent of its internal structure and composition. (ii) The outcome of any non-gravitational experiments is independent of the velocity of the freely falling reference frame in which it is performed. (iii) The outcome of any non-gravitational experiments is independent of where and when in the universe it is performed. There exist many local experiments in solar-system
supporting EEP, which implies the metric theories of gravity. In a metric theory of gravity, spacetime is endowed with a metric, and the trajectories of freely falling "test" bodies are geodesics of that metric. Moreover, in local freely falling reference frames, the non-gravitational laws of physics are those written in the language of Special Relativity.
Actually, besides GR, both $f(\mathcal{R})$ theories and BDT belong to metric theories of gravity. For metric theories, gravity is still geometry with diffeomorphism invariance as in GR.
The differences between them are just reflected in dynamical equations and additional variables.
Hence, a background-independent and non-perturbative quantization for metric theories of
gravity is preferable. Since metric $f(\mathcal{R})$ theories, as well as BDT, are a class of representative metric theories, which
have been received most attention, we will take them as examples to carry out the extension of LQG to metric theories.
Throughout the paper, we use Greek alphabet for spacetime indices,
Latin alphabet a,b,c,..., for spatial indices, and i,j,k,..., for
internal indices.

\section{Connection Dynamics of $f(\mathcal{R})$ and Brans-Dicke Theories }

By introducing an
independent variable $s$ and a Lagrange multiplier $\phi$, an
action equivalent to (\ref{action0}) of $f(\mathcal{R})$ theories is proposed as
\be
S[g,\phi,s]=\frac12\int d^4x\sqrt{-g}(f(s)-\phi(s-\mathcal
{R})).\label{action}
\ee
The variation of (\ref{action}) with
respect to $s$ yields $\phi=df(s)/{ds}\equiv f'(s)$. Assuming $s$ could be resolved from the above equation, action
(\ref{action}) is reduced to
\be
S[g,\phi]=\frac12\int
d^4x\sqrt{-g}(\phi \mathcal {R}-\xi(\phi))\equiv\int d^4x \mathcal {L}
\label{action1}
\ee
where $\xi(\phi)\equiv\phi s-f(s)$. Comparing action (\ref{action1}) with (\ref{BD}), it is obvious that metric $f(\mathcal{R})$ theories could be regarded as a particular kind of generalized BDT with certain potentials of $\phi$ and a vanishing coupling parameter $\omega$. The virtue
of (\ref{action1}) is that it admits a treatable Hamiltonian
analysis \cite{Na}. By doing 3+1 decomposition and Legendre
transformation, the Hamiltonian of metric $f(\mathcal{R})$ gravity
can be derived as a liner combination of first-class constraints as:
\be
H_{total}=\int_\Sigma N^aV_a+NH,\nonumber
\ee
where $N$ and $N^a$ are
the lapse function and shift vector respectively, and the
diffeomorphism and Hamiltonian constraints read
\ba
V_a
&=&-2D^b(p_{ab})+\pi\partial_a\phi,\nonumber\\
H
&=&\frac2{\sqrt{h}}(\frac{p_{ab}p^{ab}-\frac13p^2}{\phi}+\frac16\phi\pi^2-\frac13p\pi)+\frac12\sqrt{h}(\xi(\phi)-\phi R+2D_aD^a\phi),\nonumber
\ea
where $p^{ab}$ and $\pi$ are the momentum respectively conjugate to the spatial 3-metric $h_{ab}$ and scalar field $\phi$ on the spatial manifold $\Sigma$. Although the above Hamiltonian
analysis is started with the action (\ref{action1}) where a
non-minimally coupled scalar field is introduced, we can show
that the resulted Hamiltonian formalism is equivalent to the Lagrangian
formalism. The symplectic structure reads
\ba
\{h_{ab}(x),p^{cd}(y)\}=\delta^{(c}_a\delta^{d)}_b\delta^3(x,y),\ \ \ \
\{\phi(x),\pi(y)\}=\delta^3(x,y). \label{poission}
\ea
To achieve the connection dynamics of metric $f(\mathcal{R})$ modified
gravity, we let
\be
\tilde{K}^{ab}\equiv\phi
K^{ab}+\frac{h^{ab}}{2N}(\dot{\phi}-N^c\partial_c\phi),\nonumber
\ee
where $K_{ab}$ is the extrinsic curvature of $\Sigma$, and introduce
\be
E^a_i\equiv\sqrt{h}e^a_i,\ \ \ \ \tilde{K}_a^j\equiv\tilde{K}_{ab}e^{bj}, \nonumber
\ee
where $e^a_i$ is the triad such that $h_{ab}e^a_ie^b_j=\delta_{ij}$. Now we first extend the phase space of geometry to the space consisting of pairs $(E^a_i, \tilde{K}_a^i)$. It is then easy to see that the
symplectic structure (\ref{poission}) can be derived from the following
Poisson brackets:
\ba
\{E^a_j(x),E^b_k(y)\}=\{\tilde{K}_a^j(x),\tilde{K}_b^k(y)\}=0,\ \ \ \
\{\tilde{K}^j_a(x),E_k^b(y)\}=\delta^b_a\delta^j_k\delta(x,y). \ea
Thus there is a symplectic reduction from the extended phase space to the original one. Since
$\tilde{K}^{ab}=\tilde{K}^{ba}$, we have an additional constraint:
\be
G_{jk}\equiv\tilde{K}_{a[j}E^a_{k]}=0. \label{gaussian}
\ee
For second step, we make a canonical transformation by defining
\be
A^i_a=\Gamma^i_a+\gamma\tilde{K}^i_a,\nonumber
\ee
where $\Gamma^i_a$ is the
spin connection determined by $E^a_i$ and $\gamma$ is a nonzero real
number.
Then the Poisson brackets among the new variables read
\be
\{A^j_a(x),E_k^b(y)\}=\gamma\delta^b_a\delta^j_k\delta(x,y),\ \ \ \
\{A_a^i(x),A_b^j(y)\}=0.\nonumber
\ee
Now, the phase space consists of
conjugate pairs $(A_a^i,E^b_j)$ and $(\phi,\pi)$.
Combining
Eq.(\ref{gaussian}) with the compatibility condition:
$\partial_aE^a_i+\epsilon_{ijk}\Gamma^j_aE^{ak}=0$, we obtain the
standard Gaussian constraint
\be
\mathcal{G}_i=\mathcal{D}_aE^a_i\equiv\partial_aE^a_i+\epsilon_{ijk}A^j_aE^{ak},\label{gauss}
\ee
which justifies $A^i_a$ as an Ashtekar-Barbero $su(2)$-connection.
Had we let $\gamma=\pm i$, the (anti-)self-dual complex
connection formalism would be obtained.
The original diffeomorphism
constraint can be expressed in terms of new variables up to Gaussian
constraint as
\be
\mathcal{V}_a =\frac1\gamma
F^i_{ab}E^b_i+\pi\partial_a\phi,\nonumber
\ee
where
$F^i_{ab}\equiv2\partial_{[a}A^i_{b]}+\epsilon^i_{kl}A_a^kA_b^l$ is
the curvature of $A_a^i$. The original Hamiltonian constraint can be
written up to Gaussian constraint as
\ba
H_{fR}&=&\frac{\phi}{2}[F^j_{ab}-(\gamma^2+\frac{1}{\phi^2})\varepsilon_{jmn}\tilde{K}^m_a\tilde{K}^n_b]\frac{\varepsilon_{jkl}
E^a_kE^b_l}{\sqrt{h}}\nonumber\\
&+&\frac12(\frac2{3\phi}\frac{(\tilde{K}^i_aE^a_i)^2}{\sqrt{h}}+
\frac43\frac{(\tilde{K}^i_aE^a_i)\pi}{\sqrt{h}}+\frac23\frac{\pi^2\phi}{\sqrt{h}}+\sqrt{h}\xi(\phi))+\sqrt{h}D_aD^a\phi.\label{HfR}
\ea
Similar to GR, the constraints
are of first class.
Thus the total Hamiltonian is a linear combination of the constraints.
We have obtained the real $su(2)$-connection dynamical formalism of Lorentz $f(\mathcal{R})$ gravity \cite{Zh11,Zh11b}.
Note that a connection dynamical formalism of Euclidean $f(\mathcal{R})$ theories in Einstein frame was derived in \cite{Fa10}

We now turn to the BDT. By doing 3+1 decomposition and Legendre
transformation, the Hamiltonian of BDT
can be derived from the action (\ref{BD}) as a liner combination of first-class constraints as
$H_{total}=\int_\Sigma N^aV_a+NH$,
where the diffeomorphism constraint $V_a$ has the same form as that of metric $f(\mathcal{R})$ theories, while the Hamiltonian constraint reads
\cite{Zhang, Zh11c}
\ba
H&=&\frac2{\sqrt{h}}\left(\frac{p_{ab}p^{ab}-\frac12p^2}{\phi}+\frac{(p-\phi\pi)^2}{2\phi(3+2\omega)}\right)+\frac12\sqrt{h}\left(-\phi R+\frac{\omega}{\phi}(D_a\phi) D^a\phi+2D_aD^a\phi\right). \nonumber
\ea
In comparison with metric $f(\mathcal{R})$ theories, there is a new kinetic term $\frac{\omega\sqrt{h}}{2\phi}(D_a\phi) D^a\phi$ of the scalar field in the above Hamiltonian constraint.
However, this kinetic term will not affect the canonical transformations to connection dynamics.
Following the same canonical transformations in $f(\mathcal{R})$ theories, we obtain new conjugate variables for the geometry as:
$A^i_a=\Gamma^i_a+\gamma\tilde{K}^i_a$ and $E^a_i=\sqrt{h}e^a_i$,
as well as the Gaussian constraint (\ref{gauss}).
In terms of the new variables, the Hamiltonian constraint reads
\ba
H_{BD}&=&\frac{\phi}{2}\left(F^j_{ab}-(\gamma^2+\frac{1}{\phi^2})\varepsilon_{jmn}\tilde{K}^m_a\tilde{K}^n_b\right)
\frac{\varepsilon^{jkl}
E^a_kE^b_l}{\sqrt{h}}\nonumber\\
&+&\frac1{3+2\omega}\left(\frac{(\tilde{K}^i_aE^a_i)^2}{\phi\sqrt{h}}+
2\frac{(\tilde{K}^i_aE^a_i)\pi}{\sqrt{h}}+\frac{\pi^2\phi}{\sqrt{h}}\right)+\frac{\omega}{2\phi}\sqrt{h}(D_a\phi) D^a\phi+\sqrt{h}D_aD^a\phi.\label{HBD}
\ea
Thus both $f(\mathcal{R})$ theories and BDT have been cast into connection dynamics with $su(2)$-connection as one of configuration variables.

\section{Loop Quantization of $f(\mathcal{R})$ and Brans-Dicke Theories }

Based on the connection dynamical formalisms, the nonperturbative
loop quantization procedure can be straightforwardly extended to both $f(\mathcal{R})$ theories and
BDT. Since the scalar
field in $f(\mathcal{R})$ theories and BDT still reflects gravity, it is natural to employ the
polymer-like representations for both the scalar field and the connection. The following quantum kinematical structure is valid for both $f(\mathcal{R})$ theories and BDT.

For the geometry sector, we have the unique diffeomorphism and internal gauge invariant representation for the quantum holonomy-flux algebra\footnote{See e.g. talks by Sahlmann, Giesel, Lewandowski on Loops 11 Conference in Madrid}.
The
kinematical Hilbert space of geometry reads
\be
\mathcal{H}^{geo}_{kin}=L^2(\bar{\mathcal{A}},d\mu_{AL}),\nonumber
\ee
with the so-called
spin-network basis $T_\alpha(A)\equiv T_{\alpha,j,m,n}(\bar{A})$.
The spatial
geometric operators of LQG, such as the area \cite{Ro95}, the volume \cite{As97} and the
length operators \cite{Th98,Ma10} are still valid here.
Hence, the important physical result that both the area and the volume are discrete at quantum Kinematical level remains true for loop quantum $f(\mathcal{R})$ gravity as well as loop quantum Brans-Dicke gravity.

For the polymer-like quantization of the scalar field \cite{As03}, one extends the space $\mathcal{U}$ of smooth
scalar fields to the quantum configuration space $\bar{\mathcal{U}}$.
A simple element $U\in\bar{\mathcal{U}}$ may be thought as a
point holonomy:
$U_\lambda=\exp(i\lambda\phi(x))$
at point
$x\in\Sigma$, where $\lambda$ is a real number.
By GNS
structure, there is a natural diffeomorphism invariant
measure $d\mu$ on $\bar{\mathcal{U}}$.
Thus the kinematical Hilbert
space of scalar field reads
\be
\mathcal{H}^{sca}_{kin}=L^2(\bar{\mathcal{U}},d\mu),\nonumber
\ee
with the
scalar-network basis
\be
T_X(\phi)\equiv\prod_{x_j\in X}U_{\lambda_j}(\phi(x_j)),\nonumber
\ee
where
$X=\{x_1,\dots, x_n\}$ is an arbitrary given set of finite number
of points in $\Sigma$. The total kinematical Hilbert space for
$f(\mathcal{R})$ and Brans-Dicke gravity reads $\mathcal{H}_{kin}:=\mathcal{H}^{geo}_{kin}\otimes
\mathcal{H}^{sca}_{kin}$, with an orthonormal basis
\be
T_{\alpha,X}(A,\phi)\equiv T_{\alpha}(A)\otimes T_{X}(\phi).\nonumber
\ee
The holonomy $h_e(A)=\mathcal {P}\exp\int_eA_a$, flux $E(S,f):=\int_S\epsilon_{abc}E^a_if^i$, point holonomy $U_\lambda$, and smeared momentum $\pi(R):=\int_R
d^3x\pi(x)$ of scalar field become basic operators in $\mathcal{H}_{kin}$.
Their actions read respectively
\ba
\hat{h}_e(A)\Psi(A,\phi)&=&h_e(A)\Psi(A,\phi),\ \ \ \
\hat{E}(S,f)\Psi(A,\phi)=i\hbar\{E(S,f),\Psi(A,\phi)\},\nonumber\\
\hat{U}_\lambda(\phi(x))\Psi(A,\phi)&=&\exp(i\lambda\phi(x))\Psi(A,\phi),\ \ \ \
\hat{\pi}(R)\Psi(A,\phi)=i\hbar\{\pi(R),\Psi(A,\phi)\}.\nonumber
\ea
As in
LQG, it is straightforward to promote the
Gaussian constraint $\mathcal {G}(\Lambda)$ to a well-defined
operator. The kernel of $\hat{\mathcal{G}}(\Lambda)$ in $\mathcal{H}_{kin}$ is the internal gauge invariant
Hilbert space $\mathcal {H}_G$, with gauge invariant spin-scalar-network
basis:
$T_{s,c}(A,\phi)\equiv T_{s=(\alpha,j,i)}(A)\otimes T_{X}(\phi)$.
Since the diffeomorphisms of $\Sigma$ act covariantly
on the cylindrical functions in $\mathcal {H}_G$, the so-called
group averaging technique can be employed to solve the
diffeomorphism constraint \cite{Th07,As04,Ma07}.
Thus we can also obtain the desired diffeomorphism and gauge invariant Hilbert space $\mathcal{H}_{Diff}$ for $f(\mathcal{R})$ and Brans-Dicke gravity.

The quantum dynamics is a nontrivial issue. To implement the Hamiltonian constraint
at quantum level, we need to deal with the metric $f(\mathcal{R})$ and Brans-Dicke theories separately. The smeared version of the Hamiltonian constraint (\ref{HfR}) of metric $f(\mathcal{R})$ theories can be written in regular order as:
$H_{fR}(N)=\sum^7_{i=1}H_i(N)$. Comparing it with that of GR in connection formalism, the new
ingredients that we have to deal with are
$\phi(x),\phi^{-1}(x),\xi(\phi)$ and the four terms: $H_3(N), H_4(N), H_5(N), H_7(N)$. By introducing
certain small constant $\lambda_0$, an operator corresponding to the
scalar $\phi(x)$ can be defined as
\be
\hat{\phi}(x)=\frac{1}{2i\lambda_0}(U_{\lambda_0}(\phi(x))-U_{-\lambda_0}(\phi(x))).\nonumber
\ee
The ambiguity of $\lambda_0$ is the price that we have to pay in
order to represent field $\phi$ in the polymer-like representation.
To further define an operator corresponding to $\phi^{-1}(x)$, we
can use the classical identity
\be
\phi^{-1}(x)=sgn[\phi]\left(l^{-1} sgn[\phi]\{|\phi(x)|^{l},\pi(R)\}\right)^{\frac{1}{1-l}},\label{phi-1}
 \ee
 for any rational number $l\in(0,1)$.
For example, one may choose $l=\frac12$ for positive $\phi(x)$ and
replace the Poisson bracket by commutator to define
\be
\hat{\phi}^{-1}(x)=(\frac{2}{i\hbar}[\sqrt{\hat{\phi}(x)},\hat{\pi}(R)])^2.\nonumber
\ee
Then all the functions $\xi(\phi)$ which can be expanded as powers
of $\phi(x)$ have been quantized.
For other non-trivial types of
$\xi(\phi)$, we may replace the argument $\phi$ by $\hat{\phi}$, provided that no divergence would arise after the
replacement.
In the case where divergence does appear, there remain
the possibilities to employ tricks similar to Eq.(\ref{phi-1}) to
deal with it.
Hence it is reasonable to believe that most physically
interesting functions $\xi(\phi)$ can be quantized.
Then it is
straight-forward to quantize the term $H_6(N)$ in the Hamiltonian constraint as an operator acting on an basis vector $T_{\alpha,X}$ as
\be
\hat{H}_6(N)\cdot T_{\alpha,X} =\frac12\sum_{v\in
V(\alpha)}N(v)\hat{\xi}(\phi(v))\hat{V_v}\cdot T_{\alpha,X}.\nonumber
\ee
Moreover, by the
regularization techniques developed for the Hamiltonian constraint
operators of LQG and polymer-like scalar
field \cite{Th07,Ma06}, all the terms $H_3(N),H_4(N),H_5(N),H_7(N)$ can be
quantized as operators acting on cylindrical functions in
$\mathcal{H}_{kin}$ in state-dependent ways.
The regularization procedure involves re-expressing some variables by Thiemann's trick, point-splitting, triangulating $\Sigma$ in
adaptation to some graph $\alpha\cup X$ underling a cylindrical function, and replacing connections by holonomies.
Replacing the classical variables and Poisson brackets by the corresponding operators and commutators respectively, we can obtain the regulated Hamitonian constraint operator $\hat{H}_{fR}^\epsilon(N)$.
It turns out that the regulators of $\hat{H}_3^\epsilon(N),\hat{H}_4^\epsilon(N),\hat{H}_5^\epsilon(N)$ can be straightforwardly removed. However, the action of $\hat{H}^\epsilon_7(N)$ on a basis vector $T_{\alpha,X}$ is graph changing.
It adds a finite number of vertices with representation $\lambda_0$
at $t(s_I(v))=\epsilon$ for edges $e_I(t)$ starting from each high($\geq 3$)-valent
vertex of $\alpha$.
The family of operators
$\hat{H}^\epsilon_7(N)$ fails to be weakly convergent
when $\epsilon\rightarrow 0$.
However, due to the diffeomorphism
covariant properties of the triangulation, the limit operator can be
well defined via the so-called uniform Rovelli-Smolin topology
induced by diffeomorphism-invariant states.
The operators corresponding to the four new terms in the Hamiltonian constraint act on a basis vector as follows.
\ba \hat{H}_3(N)\cdot T_{\alpha,X} &=&\frac{4}{3\gamma^3(i\hbar)^4}
\sum_{v\in V(\alpha)}N(v)
\hat{\phi}^{-1}(v)
[\hat{H}^{Euc}(1),\sqrt{\hat{V}_v}]\ [\hat{H}^{Euc}(1),\sqrt{\hat{V}_{v}}]\cdot T_{\alpha,X}, \nonumber
\ea
\ba
\hat{H}_4(N)\cdot T_{\alpha,X} &=&-\sum_{v\in
V(\alpha)\cap X}\frac{2^{20}N(v)}{3^6\gamma^6(i\hbar)^6 E(v)^2} \hat{\pi}(v)
\sum_{v(\Delta)=v(\Delta')=v} \varepsilon(s_L,s_M,s_N)\varepsilon(s_I,s_J,s_K)\epsilon^{LMN} \nonumber\\
&\times& Tr(\tau_i\hat{h}_{s_L(\Delta)}[\hat{h}^{-1}_{s_L(\Delta)},\hat{\tilde{K}}]) Tr(\tau_i\hat{h}_{s_M(\Delta)}[\hat{h}^{-1}_{s_M(\Delta)},(\hat{V}_{v})^{3/4}]
\hat{h}_{s_N(\Delta)}[\hat{h}^{-1}_{s_N(\Delta)},(\hat{V}_{v})^{3/4}])\epsilon^{IJK} \nonumber\\
&\times&
Tr(\hat{h}_{s_I(\Delta')}[\hat{h}^{-1}_{s_I(\Delta')},(\hat{V}_{v})^{1/2}]
\hat{h}_{s_J(\Delta')}[\hat{h}^{-1}_{s_J(\Delta')},(\hat{V}_{v})^{1/2}] \hat{h}_{s_K(\Delta')}[\hat{h}^{-1}_{s_K(\Delta')},(\hat{V}_{v})^{1/2}])\cdot
T_{\alpha,X}, \nonumber
\ea
\ba
\hat{H}_5(N)\cdot T_{\alpha,X} &=&\sum_{v\in
V(\alpha)\cap X}\frac{2^{18}N(v)}{3^5\gamma^6(i\hbar)^6 E(v)^2}
\hat{\pi}(v)\phi(v)\hat{\pi}(v)\sum_{v(\Delta)=v(\Delta')=v}\varepsilon(s_I,s_J,s_K)\varepsilon(s_L,s_M,s_N)
\nonumber\\
&\times&\epsilon^{IJK} Tr(\hat{h}_{s_I(\Delta)}[\hat{h}^{-1}_{s_I(\Delta)},(\hat{V}_{v})^{1/2}]\hat{h}_{s_J(\Delta)}[\hat{h}^{-1}_{s_J(\Delta)},(\hat{V}_{v})^{1/2}] \hat{h}_{s_K(\Delta)}[\hat{h}^{-1}_{s_K(\Delta)},(\hat{V}_{v})^{1/2}])\epsilon^{LMN}\nonumber\\
&\times&Tr(\hat{h}_{s_L(\Delta')}[\hat{h}^{-1}_{s_L(\Delta')},(\hat{V}_{v})^{1/2}]
\hat{h}_{s_M(\Delta')}[\hat{h}^{-1}_{s_M(\Delta')},(\hat{V}_{v})^{1/2}]\hat{h}_{s_N(\Delta')}[\hat{h}^{-1}_{s_N(\Delta')},(\hat{V}_{v})^{1/2}])\cdot T_{\alpha,X}, \nonumber
\ea
\ba
\hat{H}_7(N)\cdot T_{\alpha,X} &=&-\sum_{v\in
V(\alpha)}\frac{2^{7}N(v)}{3\gamma^2i\lambda_0(i\hbar)^2 E(v)}\sum_{b(e)=v}\hat{X}^i_e\sum_{v(\Delta)=v}\varepsilon(s_I,s_J,s_K)\nonumber\\
&\times&\epsilon^{IJK}\hat{U}^{-1}_{\lambda_0}(\phi(s_{s_I(\Delta')}))
\left(\hat{U}_{\lambda_0}(\phi(t_{s_I(\Delta')}))-\hat{U}_{\lambda_0}(\phi(s_{s_I(\Delta')}))\right)\nonumber\\
&\times&Tr(\tau_i\hat{h}_{s_J(\Delta)}[\hat{h}^{-1}_{s_J(\Delta)},(\hat{V}_{v})^{1/2}]
\hat{h}_{s_K(\Delta)}[\hat{h}^{-1}_{s_K(\Delta)},(\hat{V}_{v})^{1/2}])\cdot T_{\alpha,X},\nonumber
\ea
where $\varepsilon(s_I, s_J,
s_K):=\mathrm{sgn}(\det(\dot{s}_I\dot{s}_J\dot{s}_K)(v))$ takes
the values $(+1, -1, 0)$ if the tangents of the three segments
$s_I,s_J,s_K$ at $v$ (in that sequence) form a matrix of positive,
negative or vanishing determinant, and $E(v)\equiv\Big({n(v)\atop 3}\Big)$ with $n(v)$ denoting the valence of the vertex $v$. The whole Hamiltonian constraint operator $\hat{H}(N)=\sum_v\hat{H}(N)_v$ is internal gauge invariant and hence also well defined
in the gauge invariant Hilbert space $\mathcal {H}_G$.

Although $\hat{H}(N)$ can dually act on the
diffeomorphism invariant states, there is no guarantee for the
resulted states to be still diffeomorphism invariant.
Hence it is
difficult to define a Hamiltonian constraint operator directly on the diffeomorphism invariant Hilbert space
$\mathcal{H}_{Diff}$.
The idea of resolution is to introduce the master constraint as in LQG \cite{Th03}:
\begin{eqnarray}
\textbf{M}:=\frac{1}{2}\int_\Sigma
d^3x\frac{|H(x)|^2}{\sqrt{|\det h(x)|}}.\nonumber
\end{eqnarray}
One then gets
the master constraint algebra as a Lie algebra:
\begin{eqnarray}
\{\mathcal{V}(\vec{N}),\ \mathcal{V}(\vec{N}')\}&=&\mathcal{V}([\vec{N},\vec{N}']),\nonumber\\
\{\mathcal{V}(\vec{N}),\ \textbf{M}\}&=&0,\nonumber\\
\{\textbf{M},\ \textbf{M}\}&=&0,\nonumber
\end{eqnarray}
where the subalgebra of diffeomorphism constraints forms an ideal.
So it is possible to define a corresponding master constraint
operator on $\mathcal{H}_{Diff}$. The master constraint can be regulated via a point-splitting
strategy as \cite{HM}:
\ba \mathcal {M}^\epsilon=\frac12\int_\Sigma d^3y\int_\Sigma
d^3x\chi_\epsilon(x-y)\frac{{H(x)}}{\sqrt{V_{U^\epsilon_x}}}\frac{{H(y)}}{\sqrt{V_{U^\epsilon_y}}}.\nonumber
\ea
Introducing a partition $\mathcal {P}$ of the 3-manifold
$\Sigma$ into cells $C$, we have an operator
$\hat{H}^\varepsilon_{C,\alpha}$ acting on gauge invariant spin-scalar-network basis
$T_{s,c}$ via a state-dependent triangulation as
\ba
\hat{H}^\varepsilon_{C,\alpha}\cdot T_{s,c}\equiv(\hat{H}^\varepsilon_{C,GR}+\sum^7_{i=3}\hat{H}^\varepsilon_{C,i})\cdot T_{s,c}=\sum_{v\in
V(\alpha)}\chi_C(v)(\sum_{v(\Delta)=v}\hat{H}^{\varepsilon,\Delta}_{GR,v}+\sum^7_{i=3}
\hat{H}^\varepsilon_{i,v}) \cdot
T_{s,c},\nonumber
\ea
where $\alpha$ denotes the underlying
graph of the spin-network state $T_s$. Here,
\ba
\hat{H}^{\varepsilon,\Delta}_{GR,v}&=&\frac{32}{3E(v)}\varepsilon(s_I,s_J,s_K)\epsilon^{IJK}\big(\frac{1}{i\hbar\gamma}\hat{\phi}(v)
Tr(\hat{h}_{\alpha_{IJ}(\Delta)}\hat{h}_{s_K(\Delta)}
[\hat{h}^{-1}_{s_K(\Delta)},\sqrt{\hat{V}_{U^\epsilon_{v}}}])\nonumber\\
&-&\frac{1}{(i\hbar)^3\gamma^3}(\hat{\phi}^{-1}(v)+\gamma^2\hat{\phi}(v))
Tr(\hat{h}_{s_I(\Delta)}[\hat{h}^{-1}_{s_I(\Delta)},\hat{\tilde{K}}]\hat{h}_{s_J(\Delta)}[\hat{h}^{-1}_{s_J(\Delta)},\hat{\tilde{K}}]\hat{h}_{s_K(\Delta)}
[\hat{h}^{-1}_{s_K(\Delta)},\sqrt{\hat{V}_{U^\epsilon_{v}}}])\big),\nonumber
\ea
\ba
\hat{H}_{3,v}^\varepsilon =
\frac{16}{3\gamma^3(i\hbar)^4}
\hat{\phi}^{-1}(v)
[\hat{H}_{Euc}(1),(\hat{V}_{U^\epsilon_{v}})^{1/4}][\hat{H}_{Euc}(1),(\hat{V}_{U^\epsilon_{v}})^{1/4}], \nonumber
\ea
and
$\hat{H}^\varepsilon_{4,v},\
\hat{H}^\varepsilon_{5,v},\
\hat{H}^\varepsilon_{6,v},\
\hat{H}^\varepsilon_{7,v}$ can be defined similarly \cite{Zh11b}.
The family of
operators $\hat{H}^\varepsilon_{C}$ are cylindrically
consistent up to diffeomorphisms.
The inductive limit operator $\hat{H}_{C}$ can be well defined by the uniform Rovelli-Smolin topology. Thus we can define master constraint operator $\hat{\mathcal {M}}$
on diffeomorphism invariant states as
\ba
(\hat{\mathcal {M}}\Phi_{Diff})T_{s,c}=\lim_{\mathcal{P}\rightarrow\Sigma;\varepsilon,\varepsilon'\rightarrow
0}\Phi_{Diff}[\frac12\sum_{C\in\mathcal
{P}}\hat{H}^\varepsilon_C(\hat{H}^{\varepsilon'}_C)^\dagger T_{s,c}].\nonumber
\ea
It is obvious that $\hat{\mathcal {M}}$ is diffeomorphism invariant.
For any given diffeomorphism invariant spin-scalar-network
state $T_{[s,c]}$, the norm $\|\hat{\mathcal {M}}T_{[s,c]}\|_{Diff}$ is finite.
So $\hat{\mathcal {M}}$ is densely defined in $\mathcal{H}_{Diff}$.
Moreover, $\hat{\mathcal {M}}$ is positive and symmetric and hence admits a unique self-adjoint Friedrichs
extension.
It is then possible to obtain the physical
Hilbert space of quantum $f(\mathcal{R})$ gravity by the direct integral
decomposition of $\mathcal{H}_{Diff}$ with respect to $\hat{\mathcal{M}}$.

Now we turn to the quantum dynamics of BDT. Comparing the Hamiltonian
constraint (\ref{BD}) of Brans-Dicke gravity with that of metric $f(\mathcal{R})$ gravity in connection formalism, the only new
ingredient that we have to deal with is the kinetic term:
\be
H_k(N)=\int_\Sigma d^3xN\frac{\omega}{2\phi}\sqrt{h}(D_a\phi) D^a\phi.\nonumber
\ee
By the regularization techniques similar to those for $f(\mathcal{R})$ gravity, it can be quantized as a well-defined operator in $\mathcal{H}_{kin}$ as \cite{Zhang}
\ba
\hat{H}_k\cdot T_{\alpha,X} &=&\sum_{v\in
V(\alpha)}\frac{2^{17}\omega N(v)}{3^6\gamma^4(i\lambda_0)^2(i\hbar)^4E(v)^2}\hat{\phi}^{-1}(v)
\sum_{v(\Delta)=v(\Delta')=v}\varepsilon(s_I,s_J,s_K)
\varepsilon(s_L,s_M,s_N)\nonumber\\
&\times&\epsilon^{LMN}\hat{U}^{-1}_{\lambda_0}(\phi(s_{s_L(\Delta)}))
[\hat{U}_{\lambda_0}(\phi(t_{s_L(\Delta)}))-\hat{U}_{\lambda_0}(\phi(s_{s_L(\Delta)}))]\nonumber\\
&\times&Tr(\tau_i\hat{h}_{s_M(\Delta)}[\hat{h}^{-1}_{s_M(\Delta)},(\hat{V}_v)^{3/4}]
\hat{h}_{s_N(\Delta)}[\hat{h}^{-1}_{s_N(\Delta)},(\hat{V}_v)^{3/4}]) \nonumber\\
&\times& \epsilon^{IJK}\hat{U}^{-1}_{\lambda_0}(\phi(s_{s_I(\Delta')}))
[\hat{U}_{\lambda_0}(\phi(t_{s_I(\Delta')}))-\hat{U}_{\lambda_0}(\phi(s_{s_I(\Delta')}))]\nonumber\\
&\times&Tr(\tau_i\hat{h}_{s_J(\Delta')}[\hat{h}^{-1}_{s_J(\Delta')},(\hat{V}_{v})^{3/4}]
\hat{h}_{s_K(\Delta')}[\hat{h}^{-1}_{s_K(\Delta')},(\hat{V}_{v})^{3/4}])\cdot
T_{\alpha,X}. \nonumber
\ea
Since all the other terms in the Hamiltonian constraint are equal to the corresponding terms of metric $f(\mathcal{R})$ theories up to coefficients, the whole Hamiltonian constraint operator of BDT has been well defined. For the same reason as in $f(\mathcal{R})$ gravity, we also wish to define a master constraint operator for Brans-Dicke gravity.
The "square root" of the new term in the regulated master constraint can be quantized as
\ba
\hat{H}^\varepsilon_{k,v} &=&\sum_{v(\Delta)=v(\Delta')=v}\frac{2^{15}\omega}{3^4\gamma^4(i\lambda_0)^2(i\hbar)^4E(v)^2}
\varepsilon(s_I,s_J,s_K)
\varepsilon(s_L,s_M,s_N)\nonumber\\
&\times&\hat{\phi}^{-1}(v)\epsilon^{LMN}\hat{U}^{-1}_{\lambda_0}(\phi(s_{s_L(\Delta)}))
[\hat{U}_{\lambda_0}(\phi(t_{s_L(\Delta)}))-\hat{U}_{\lambda_0}(\phi(s_{s_L(\Delta)}))]\nonumber\\
&\times&Tr(\tau_i\hat{h}_{s_M(\Delta)}[\hat{h}^{-1}_{s_M(\Delta)},(\hat{V}_v)^{1/2}]
\hat{h}_{s_N(\Delta)}[\hat{h}^{-1}_{s_N(\Delta)},(\hat{V}_v)^{3/4}]) \nonumber\\
&\times&\epsilon^{IJK}\hat{U}^{-1}_{\lambda_0}(\phi(s_{s_I(\Delta')}))
[\hat{U}_{\lambda_0}(\phi(t_{s_I(\Delta')}))-\hat{U}_{\lambda_0}(\phi(s_{s_I(\Delta')}))]\nonumber\\
&\times&Tr(\tau_i\hat{h}_{s_J(\Delta')}[\hat{h}^{-1}_{s_J(\Delta')},(\hat{V}_{v})^{1/2}]
\hat{h}_{s_K(\Delta')}[\hat{h}^{-1}_{s_K(\Delta')},(\hat{V}_{v})^{3/4}]), \nonumber
\ea
from which a positive and self-adjoint master constraint operator can be defined on $\mathcal{H}_{Diff}$. Thus the master constraint program is also valid for BDT.

\section{Summary and Outlook}

Based on the above loop quantization procedure of metric $f(\mathcal{R})$ theories and BDT, we can propose a general quantization scheme for 4-dimensional metric theories of gravity. The prerequisite is that the theories should have well-defined geometrical dynamics, which means a Hamiltonian formalism with 3-metric $h_{ab}$ as one of configuration variables and a closed (first-class) constraint algebra (perhaps after solving some second-class constraints). Without loss of generality, we assume that the classical phase space of the theory consists of
conjugate pairs $(h_{ab},p^{ab})$ and $(\phi_A,\pi^A)$, where $\phi_A$ could be a rather arbitrary scalar, vector, tensor or spinor field. Then the quantization scheme consists of the following steps.
(i) To obtain a connection dynamical formalism, we first enlarge the phase space by transforming to the
triad formulation as
\ba
(h_{ab},p^{ab})\rightsquigarrow (E^a_j\equiv\sqrt{h}h^{ab}e_{aj},\tilde{K}_a^j\equiv\tilde{K}_{ab}e^{bj}),\nonumber
\ea
where $\tilde{K}_{ab}=\frac{2}{\sqrt{h}}(p_{ab}-\frac12ph_{ab})$ and $\tilde{K}_{a[i} E^a_{j]}=0$.
Then we make a canonical transformation to connection formulation as:
\ba
(E^a_j, \tilde{K}_a^j)\rightsquigarrow (E^a_j, A_a^j\equiv \Gamma_a^j+\gamma \tilde{K}_a^j),\nonumber
\ea
with the Gaussian constraint, $\mathcal{D}_aE^a_i\equiv\partial_aE^a_i+\epsilon_{ijk}A^j_aE^{ak}=0$, appeared by construction. It is straightforward to write all the constraints in terms of the new variables. (ii) For loop quantization, we first find the polymer-like representation of the fields $(\phi_A,\pi^A)$, together with the LQG representation of the holonomy-flux algebra.
Then the kinematical Hilbert space reads $\mathcal{H}_{kin}:=\mathcal{H}^{geo}_{kin}\otimes
\mathcal{H}^{\phi}_{kin}$, where the basic operators and geometrical operators could be well defined.
By implementing the Gaussian and diffeomorphism constraints as in standard LQG, we could get the gauge and diffeomorphism invariant Hilbert space as: $\mathcal{H}_{kin}\rightsquigarrow\mathcal{H}_{G}\rightsquigarrow\mathcal{H}_{Diff}$.
To implement quantum dynamics, one may first construct the Hamiltonian constraint operator at least in $\mathcal{H}_{G}$.
Usually it could not be well defined on $\mathcal{H}_{Diff}$. Then we can construct master constraint operator in $\mathcal{H}_{Diff}$ by using the structure of the Hamiltonian operator.
(iii) One may try to understand the physical Hilbert space by the direct integral
decomposition of $\mathcal{H}_{Diff}$ with respect to the master constraint operator.
(iv) One may also do certain semiclassical analysis in order to confirm the classical limits of the Hamiltonian and master constraint operators as well as the constraint algebra. The low energy physics is also expected in the analysis.
(v) Finally, to complement above canonical approach, we can also try the covariant path integral (spin foam) quantization. It should be noted that in the present work we only finished steps (i) and (ii) for metric $f(\mathcal{R})$ theories and BDT.

To summarize, our main results in two folds. First, the 4-dimensional connection dynamics of metric
$f(\mathcal{R})$ theories, as well as Brans-Dicke theory, have been obtained by canonical transformations from
their geometrical dynamics.
Thus GR is not the unique theory of gravity with connection dynamical character.
Second, due to the $su(2)$-connection
dynamical formalism, the 4-dimensional metric
$f(\mathcal{R})$ theories and Brans-Dicke theory have been nonperturbatively quantized by extending LQG scheme.
Hence, the non-perturbative loop quantization
procedure is not only valid for GR but also valid for a
general class of metric theories of gravity. In fact, it is not difficult to extend LQG further to general scalar-tensor theories of gravity \cite{Zh11c}. Moreover, since higher dimensional scalar-tensor theories of gravity have well-defined Hamiltonian geometrical dynamics, and the symplectic reduction of Bodendorfer-Thiemann-Thurn connection formalism to metric formalism does not depend on dynamics, LQG may also be extended to higher($>4$) dimensional scalar-tensor theories of gravity \cite{HZM}.

Of course, there are still many open issues on the extension of LQG. It is desirable to find suitable actions for
the connection dynamics of $f(\mathcal{R})$ theories and that of Brans-Dicke theory. We will explore the applications of loop quantum $f(\mathcal{R})$ and Brans-Dicke theories to cosmology and black holes in future work \cite{GWZM}.
It is also desirable to quantize metric theories of gravity by the covariant spin foam approach \cite{Zhou}.
To conclude, our conservative observation is that LQG could be applicable to metric theories of gravity (with well-defined geometrical dynamics) in arbitrary dimensions. However, a caution arises from the difference between the 4(or 3)-dimensional and higher-dimensional connection dynamical formulations of GR. In particular, in 4-dimensional case, we have both Ashtekar-Barbero connection dynamics and Bodendorfer-Thiemann-Thurn connection dynamics of GR.
Are the quantum theories corresponding to them unitary equivalent to each other \footnote{See the talk by Thomas Thiemann on Loops 11 Conference in Madrid}?
If the answer is negative, are there any theoretical criteria for judging them?
Is Bodendorfer-Thiemann-Thurn connection dynamics only preferable in higher dimensions? Another interesting question is whether LQG can be extended to non-metric theories, e.g., metric-affine $f(\mathcal{R})$ theories. All these open issues are fascinating and deserve future investigating.

\section*{Acknowledgments} 
The author would like to thank Thomas Thiemann and Xiangdong Zhang for helpful discussion and fruitful collaboration and acknowledge the organizers of Loops 11 conference for the
financial support. This work is supported in part by NSFC
(Grant No.10975017) and the Fundamental Research Funds for the
Central Universities.

\end{document}